\documentclass[twocolumn,prl]{revtex4}
\usepackage{epsfig,amssymb,amsmath}

\usepackage{graphicx}

\usepackage{epstopdf}

\usepackage{epsfig}

\usepackage{amsmath}
\usepackage[T1]{fontenc}
\usepackage{float}
\usepackage{dcolumn}
\usepackage{hyperref}
\usepackage{amsthm}
\usepackage{amssymb}

\begin{document}
\title{The ac Driven Annular Josephson Junctions: The Missing Steps
}
\author{I. R. Rahmonov$^{1,2}$, J. Teki\' c$^3$, P. Mali$^4$, and A. Irie$^5$, and Yu. M. Shukrinov$^{1,6}$,}

\address{
$^1$BLTP, JINR, Dubna, Moscow Region, 141980, Russia \\
$^2$Umarov~Physical~Technical~Institute,~TAS,~Dushanbe,~734063~Tajikistan\\
$^3$"Vin\v ca" Institute of Nuclear Sciences, Laboratory for Theoretical and Condensed Matter Physics - 020, University of Belgrade, PO Box 522, 11001 Belgrade, Serbia\\
$^4$Department of Physics, Faculty of Science, University of Novi Sad, Trg Dositeja Obradovi\' ca 4, 21000 Novi Sad, Serbia\\
$^5$Department of Electrical and Electronic Systems Engineering, Utsunomiya University, 7-1-2 Yoto, Utsunomiya 321-8585, Japan\\
$^6$Dubna State University, Dubna,  141980, Russia
}

\date{\today}

%==================================================================================================
\begin{abstract}
Examination of an annular system of underdamped Josephson junctions in the presence of external radiation showed that
the ability of the system to lock with some external radiation was determined not only by the number but also by the type of rotating excitations (fluxons or antifluxons).
Shapiro steps can be observed in the current-voltage characteristics only in the system with trapped fluxons or in the system with fluxon-antifluxon pairs.
If the trapped fluxons circulate simultaneously with fluxon-antifluxon pairs, there are no Shapiro steps regardless of the amplitude or frequency of the applied external radiation.
\end{abstract}
%==================================================================================================

\maketitle
%==================================================================

The idea that fluxon behaves as a particle-like solitary wave, which can be manipulated and controlled,
motivated creation of a new logic circuit by using Josephson fluxon as elementary bits of information~\cite{Clarke, WallNat, Fed, Herr11, Mukhanov11, Volkmann13, Takeuchi}.
In the creation of a new logic elements, particularly important are the long Josephson junctions~\cite{Mazo} described by continuous sine-Gordon equation, and the Josephson junctions parallel array by its discreet counterpart i.e. Frenkel-Kontorova model~\cite{Mazo, OBBook, ACFK, Fult, Lucci15}.
However, in long JJs, motion of fluxon strongly depends on the geometry and boundaries of the junctions,
which makes studies of fluxon dynamics very challenging.
These problems led to the creation of annular Josephson junctions~\cite{DavPRL85}, as ideal systems for the studies of fluxon dynamics,
which provide an undisturbed and tunable fluxon motion~\cite{Pfeiffer08, Ustinov93, UstPRL, Zant95, Wall, Watanabe, Nappi, Monaco}.

One of the most interesting properties of Josephson junction systems is their ability to exhibit various resonance phenomena.
In the absence of any external radiation, the so called {\it zero field steps} (ZFS)~\cite{Fult, Lomdhal, Kawamoto} appear in the current-voltage (I-V) characteristics due to resonant motion of fluxons and antifluxons inside the system.
If, on the other hand, some external radiation is applied,
the I-V characteristics exhibits the well known {\it Shapiro steps}~\cite{ShapPRL} as a
result of the locking with the external frequency.
When the locking appears at the integer values of external frequency,
the steps are called harmonics, while the locking at rational noninteger values leads to subharmonics
(for the locking at the half integer values of external frequency, the steps are called halfinteger)~\cite{ACFK}.
Though the Shapiro steps are today one of the most recognized frequency locking phenomena associated with wide variety of physical systems~\cite{ACFK}, the majority of the works~\cite{Pfeiffer08, Ustinov93, UstPRL, Zant95, Wall, Watanabe, Nappi, Monaco} on annular Josephson junctions have been focused on the resonance phenomena in the absence of external radiation.

In this study, we will examine the underdamped dynamics of an annular array of Josephson junctions (AAJJ) under the external radiation.
In contrast to previous studies of annular Josephson junctions, which were mainly focused on the case of one trapped fluxon in a small range of currents and voltages~\cite {Pfeiffer08, Ustinov93}, here, we will examine the Shapiro steps in various cases of circulating excitations (fluxons and antifluxons), in a wide range of currents and voltages in order to get the full picture of dynamical behavior.
Surprisingly, our results show that ability of the system to lock with some external radiation depends not only on the number but also on the type of excitations, i.e., whether there are only trapped fluxons or the fluxon-antifluxon pairs in the system, or the trapped fluxons circulate simultaneously with fluxon-antifluxon pairs.

\begin{figure}[htb]
 \centering
\includegraphics[width=70mm]{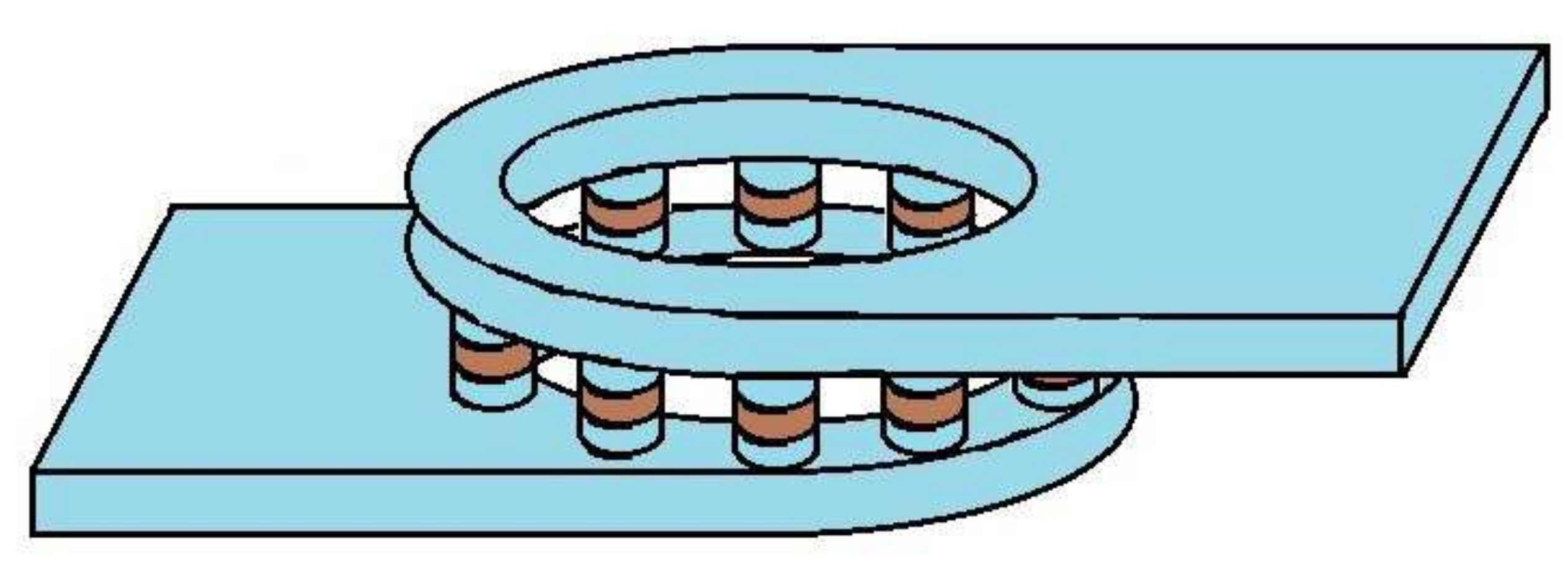}\\
\caption{(Color online). Schematic view of an annular array of Josephson junctions.
The junctions are colored in dark red.}
\label{Fig1}
\end{figure}

We consider an annular parallel array of $N$ Josephson junctions in underdamped regime presented in Fig. \ref{Fig1}.
The total length of a chain is  $L=Na$, where $a$ is a distance between the neighboring junctions.
The annular system that we are considering here can be described by the discrete version of perturbed sine-Gordon equation, which is well known as the dissipative Frenkel-Kontorova model~\cite{OBBook}:
\begin{equation}
\label{fk_equation}
\frac{d ^{2}\varphi _i}{d t^{2}}-\frac{\varphi_{i+1}+2\varphi_{i}+\varphi_{i-1}}{a^{2}}+\sin\varphi _i+\alpha\frac{d \varphi _i}{d t}=I+A\sin (\omega t),
\end{equation}
where $\varphi_{i}$ is the phase difference across the $i$-th junction, $\alpha $ is dissipation parameter, $I$ is the total or biased current through the junction, and $A$ and $\omega $ are the amplitude and frequency of external radiation, respectively.
The coupling between the neighboring junctions is described by the constant
$\frac {1}{a^2}$, where $a=\sqrt{2\pi L_{0}I_{c}/\Phi_{0}}$ is the discreteness parameter,
i.e., distance between two junctions normalized to the Josephson penetration depth.
The time is normalized with respect to the inverse plasma frequency $\omega_{p}^{-1}$, where $\omega_{p}=\sqrt{2\pi I_{c}/(\Phi_{0}C)}$,
$I_{c}$ is the critical current, $L_{0}$ and $C$ are the inductance and capacitance of single cell, respectivelly, and $\Phi_{0}=\frac {h}{2e}$ is the flux quantum~\cite{UstPD}.

In order to calculate the I-V characteristic of the AAJJ we have used the Eq. (\ref{fk_equation}) and the Josephson relation:
\begin{equation}
\label{JR_norm}
V_i=\frac{d\varphi_i}{dt}=\omega _J,
\end{equation}
where $V_i$ is the voltage of the $i$th junction normalized to $V_{0}=\hbar \omega_{p}/2e$, and
$\omega _J$ is the Josephson frequency normalized to $\omega _p$.

Our numerical simulations were performed for the periodic boundary conditions, which in discrete case have the form:
\begin{equation}
\label{bc}
\varphi_{N+1}=\varphi_{1}+2\pi M, \hspace{0.8cm} \varphi_{0}=\varphi_{N}-2\pi M,
\end{equation}
where $M$ is the number of trapped fluxons inside the system.
The spatial points $i=0$ and $i=N+1$ were assumed to be equivalent to $i=N$ and $i=1$, respectively.
We have applied the well known procedure used in Ref. \onlinecite{ShukLNCS2012} and \onlinecite{ShukJETP2012}.
The current was changed by step $\Delta I$ and for every value of $I$ the corresponding voltage $V$ was calculated,
in that way the I-V characteristic was produced.
We note that the solution at certain value of $I$ was used as the initial condition for the calculation of the next point at the value of bias current $I+\Delta I$.

\begin{figure}[h!]
\centering
\includegraphics[width=80mm]{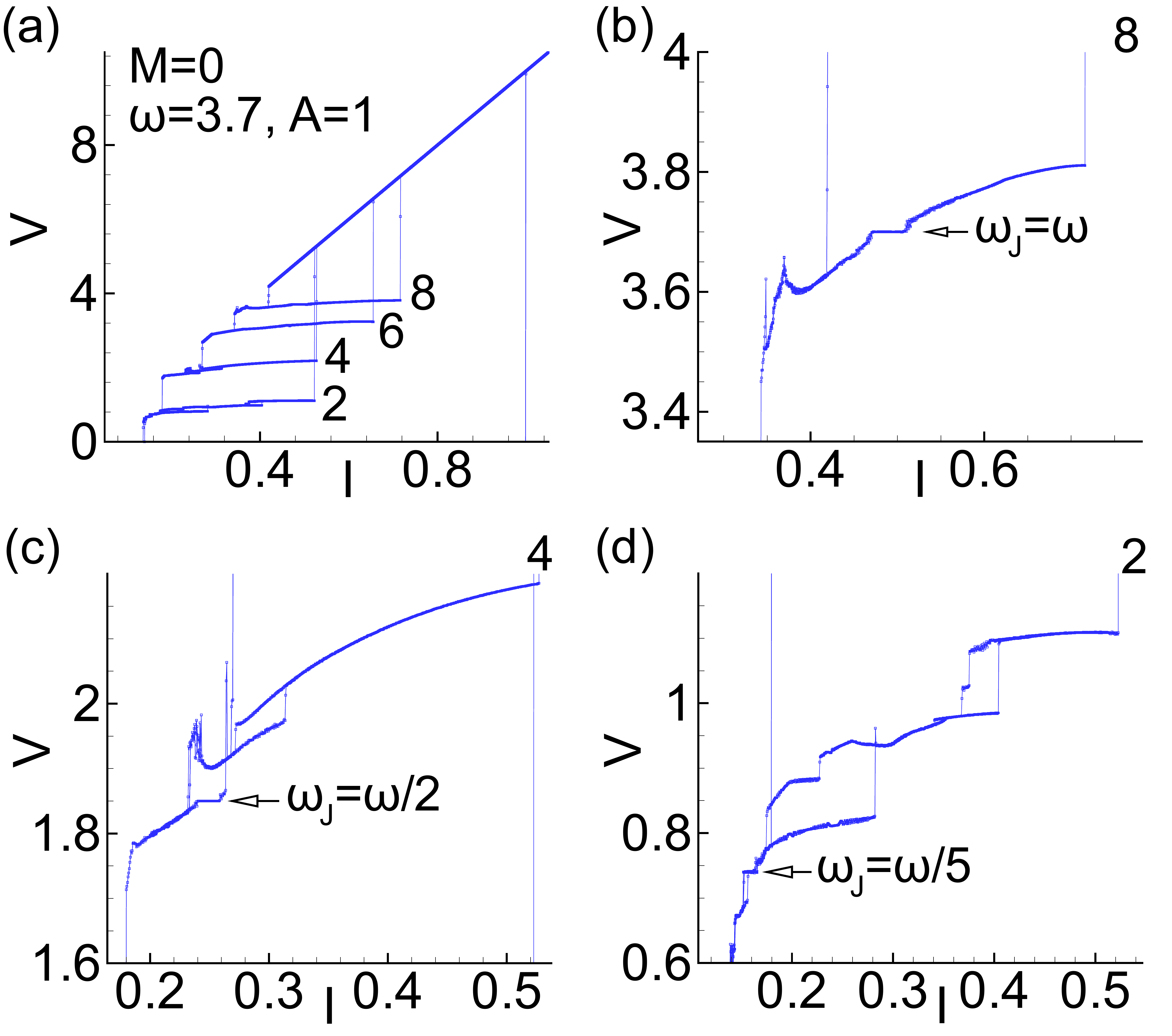}
\caption{(a) The current-voltage characteristic of the annular array of Josephson junctions for $N=10$, $a=1$, $\alpha=0.1$, and $M=0$.
The amplitude and frequency of external radiation are $A=1$ and $\omega =3.7$, respectively.
The numbers 2, 4, 6, and 8 mark the total number of fluxon and antifluxons $n$.
The zoomed parts of I-V characteristics correspond to: (b) the harmonic step on the ZFS $n=8$; (c) the halfinteger step on the ZFS $n=4$;
and (d) the subharmonic step $\frac 15$ on the ZFS  $n=2$.}
\label{Fig2}
\end{figure}

If no trapped fluxons are present in the system ($M=0$), depending on the current, one or more fluxon-antifluxon pairs are circulating along the system.
Due to the presence of external radiation, in addition to zero field steps, the system will also exhibit Shapiro steps.
In Fig. \ref{Fig2} the I-V characteristics of the AAJJ  in the presence of external radiation for $M=0$ is presented.
At first, it might look in Fig. \ref{Fig2} (a) that the I-V characteristics exhibit only four ZFSs, which correspond to $n=2,4,6$ and $8$ excitations, which appear when $V$, i.e., Josephson frequency satisfies the resonant condition $\omega_J=\frac{2\pi n u}{L}$, where $u$ is a speed of moving fluxon (antifluxon).
Here, $n=n_f+n_{af}=2n_p+M$ is the total number of excitations, i.e., fluxons and antifluxons in the system, where  $n_f$, $n_{af}$ and $n_p$ are the number of fluxons, antifluxons and fluxon-antifluxons pairs, respectively.
However, the high resolution analysis reveals also the Shapiro steps that come from the locking of Josephson frequency and the frequency of external radiation.
For a given external frequency $\omega =3.7$, the first harmonic step
appears on the $n=8$ ZFS as can be seen in Fig. \ref{Fig2} (b),
while Fig. \ref{Fig2} (c) and (d) show the halfinteger
$\frac 12\omega $ and the subharmonic step $\frac 15\omega $, which appear on the $n=4$ and $n=2$ ZFS, respectively.
We have examined the AAJJ for a wide range of applied frequencies $\omega $, and we were able to obtain Shapiro steps in the whole area of the I-V characteristics in  Fig. \ref{Fig2}.

\begin{figure}[h!]
\centering
\includegraphics[width=60mm]{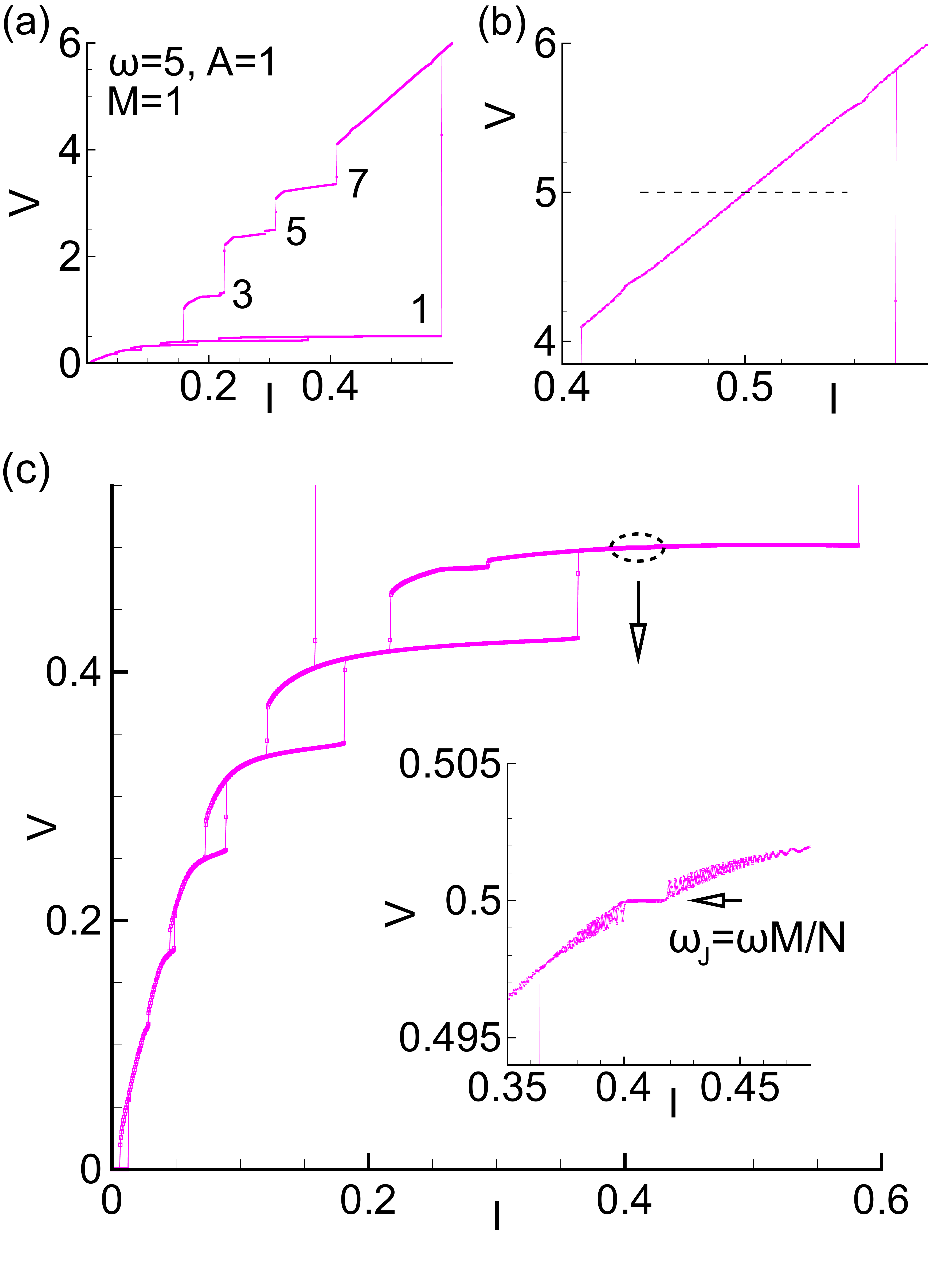}
\caption{(a) The current-voltage characteristic of the annular array of Josephson junctions for $N=10$, $a=1$, $\alpha=0.1$, $M=1$, with the amplitude and frequency of external radiation $A=1$ and $\omega =5$, respectively.
The numbers 1, 3, 5, and 7 mark the total number of fluxon and antifluxons $n$.
(b) The absence of the Shapiro step in the I-V curve, dashed line marks where the step should be.
(c) High resolution plot of the step $n=1$, which exhibits Shapiro step shown in the inset. }
\label{Fig3}
\end{figure}

If there are fluxons trapped in the system, the ability of system to exhibit Shapiro steps will completely change.
In the Fig. \ref{Fig3} the I-V characteristics of the AAJJ with one trapped fluxon ($M=1$) is presented.
In this case,
for the applied frequency of the external radiation $\omega =5$, in Fig. \ref{Fig3} (a) we would expect to see the Shapiro step at $V=5$ as well as other subharmonic steps in the I-V characteristics.
However, as we can see in  Fig. \ref{Fig3} (b) there is no Shapiro steps, and the only Shapiro step we could detect was the step $\frac {1}{10}\omega $, which appeared for $n=1$ in Fig. \ref{Fig3} (c).

\begin{figure}[h!]
\centering
\includegraphics[width=60mm]{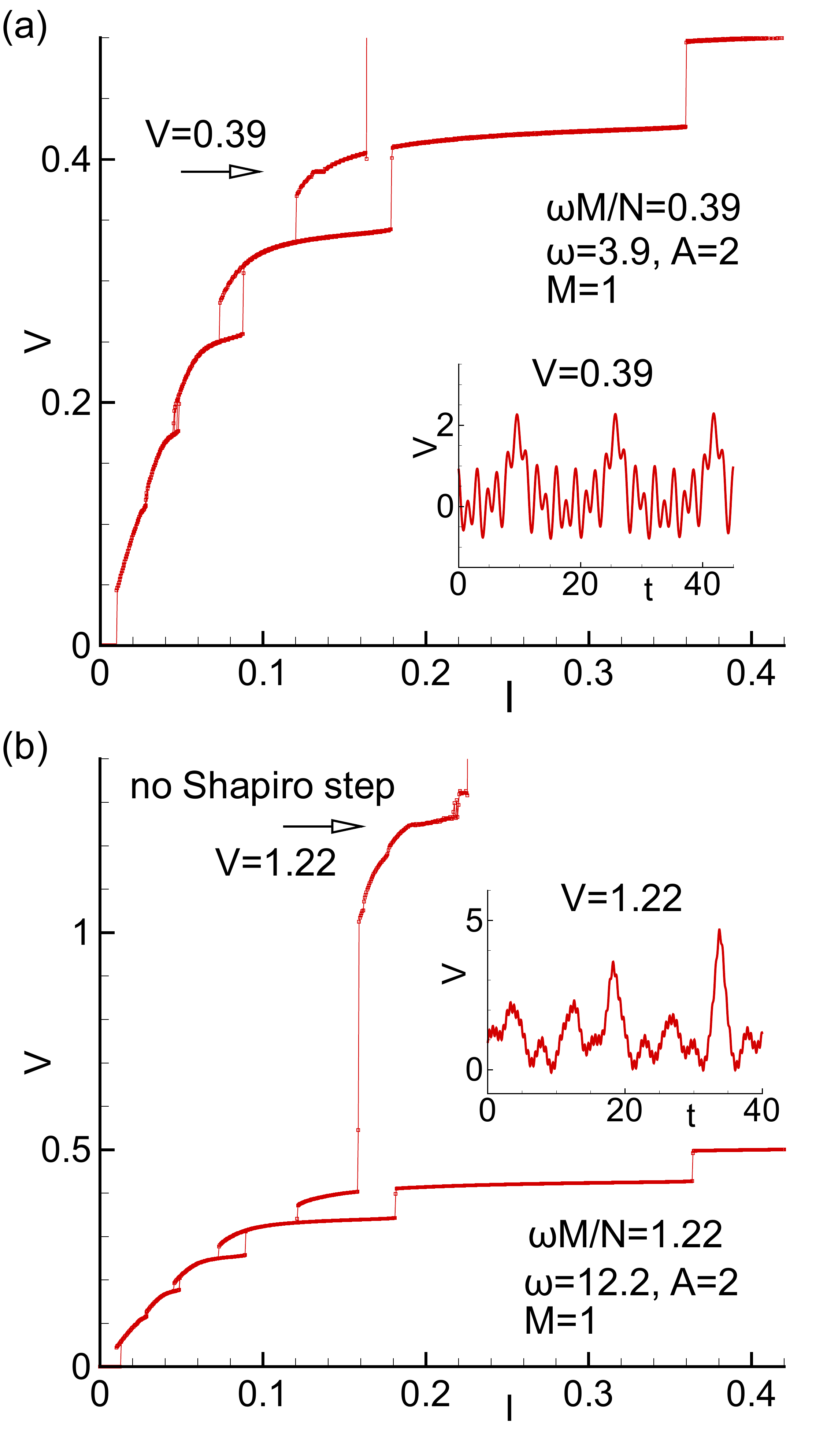}
\caption{(a) The part of the I-V characteristics corresponding to the step $n=1$ for $M=1$,
$\omega=3.9$, $A=2$.
The rest of the parameters are the same as in Fig. \ref{Fig3}.
(b) The part of the I-V characteristics corresponding to the $n=3$ ZFS for $M=1$, $\omega=12.2$, $A=2$.
Insets show the voltage time dependence at the values of $I$ marked by arrows at which Shapiro steps should appear.
}
\label{Fig4}
\end{figure}

In Fig. \ref{Fig4} two I-V characteristics at two different values of applied frequencies are presented.
As in previous case, the Shapiro step appears in Fig. \ref{Fig4} (a) at $V=0.39$, when the applied frequency is in the region of the step $n=1$, i.e. in the region where only one trapped fluxon rotates through the system.
On the other hand, if we increase the frequency to the value, which corresponds to the higher steps ($n>1$) in I-V characteristics, no Shapiro steps appear as can be seen for $V=1.22$ at the $n=3$ step in  Fig. \ref{Fig4} (b).
The voltage time dependence given in the insets
clearly shows the periodic behavior on the Shapiro step in Fig. \ref{Fig4} (a), and the nonperiodic one in Fig. \ref{Fig4} (b) in its absence.

\begin{figure}[h!]
\centering
\includegraphics[width=60mm]{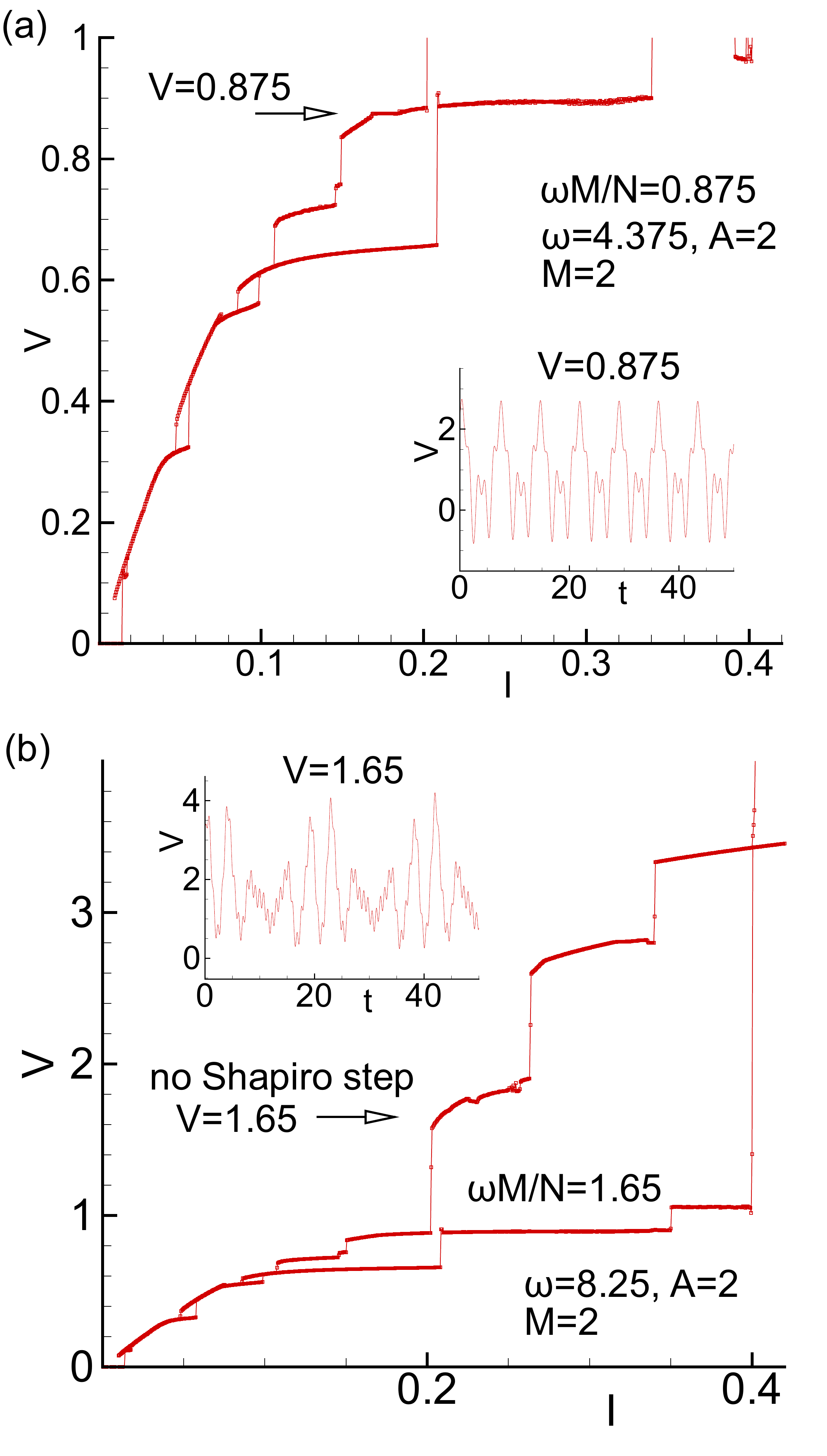}
\caption{(a) The part of the I-V characteristics corresponding to the step $n=2$ for $M=2$, $\omega=4.375$, $A=2$.
The rest of the parameters are the same as in Fig. \ref{Fig3}.
(b) The part of the I-V characteristics corresponding to the step $n=4$ for $M=2$, $\omega=8.25$, $A=2$.
Insets show the voltage time dependence at the values of $I$ marked by arrows at which Shapiro steps should appear.
}
\label{Fig5}
\end{figure}

The situation remains unchanged if more fluxons are introduced.
In Fig. \ref{Fig5}, the I-V characteristics for two trapped fluxons $M=2$ at two different values of applied frequencies are presented.
As we can see from the I-V characteristics and the corresponding voltage time dependence
in  Fig. \ref{Fig5} (a), again, Shapiro steps appear in the region, where only two fluxons are present in the system.
On the other hand, in  Fig. \ref{Fig5} (b), where $\omega $ is in the region of the $n=4$ step, i.e., in addition to two trapped fluxon there is also one fluxon-antifluxon pair, there are no Shapiro steps.

When $M=0$, we could create Shapiro steps anywhere in the I-V characteristics, however, this was not the case for $M\neq 0$.
We performed simulations for a wide range of system parameters and obtained that if there were trapped fluxons,
regardless of the value of $\omega $ or $A$,
Shapiro steps would appear only in the part of I-V characteristics
which corresponds to the step $n=M$ ($n_p=0$), where only trapped fluxons moved through the AAJJ.
If in addition to the trapped fluxons there are also fluxon-antifluxon pairs, i.e.,  $n=2n_p+M$ ($n_p\neq 0$), there would be no Shapiro steps.
This leads us to the conclusion that the appearance of Shapiro steps is somehow determined by the type of excitations, and raises the question why for $M\neq 0$ Shapiro steps do not exist if fluxons and fluxon-antifluxon pairs simultaneously circulate in the system.

In order to understand this fact let us consider first the case of one trapped fluxon in the AAJJ.
When $M=1$, in the region of the I-V characteristics, which corresponds to the $n=1$ step,
we have only one circulation fluxon, so it will pass through a junction at the equal time intervals, and consequently, this periodic motion can get locked with some external periodic radiation.
In the same way, for any $M\neq 0$ we will have $n=M$ fluxons circulating around the ring equally distributed in space and time.
As they move, they are passing through junctions in the equal time intervals and this motion can get locked with some external frequency.

However, if in addition to trapped fluxons, we have also fluxon-antifluxon pairs situation will be completely different.
Let us look at one example: the case of $n=3$ excitations present in the system.
If we have one trapped fluxon and one fluxon-antifluxon pair in which case $n=2+1=3$, this will be completely different from the case of 3 trapped fluxons, for which the total number of excitations is also $n=3$.
Three fluxons are always equally distributed in space and time and rotate passing through junctions in the same time intervals.
This will change if instead, we have two fluxons and one antifluxon.
Though they all move periodically, the antifluxon is moving in the opposite direction of the two fluxons, and so they are not any more equally distributed in space and time (distance between antifluxon and two fluxons is changing as they move).
Consequently, they will not pass through a junction in the same time intervals, but the period between two consecutive passages will constantly change, and
for that reason, it would  be impossible for the system to lock with an external radiation.
Therefore, every time there is an uneven number of fluxons and antifluxons, the system can not exhibit Shapiro steps.

In conclusion, the examination of the fluxon dynamics in an annular array of underdamped Josephson junctions demonstrated that
not only the number but also the type of rotating excitations (fluxons or antifluxons) determined
the ability of the system to lock with the external radiation.
Regardless of the amplitude or frequency of the external radiation,
the current-voltage characteristics exhibits Shapiro steps only in the system with trapped fluxons or in the system with fluxon-antifluxon pairs.
If the trapped fluxons circulate simultaneously with fluxon-antifluxon pairs, there are {\it no} Shapiro steps.
Though, this phenomenon of {\it missing steps} was observed in one particular system, the AAJJ, it could be relevant for any system where dynamics is governed by the moving fluxons and antifluxons since
any disbalance between their numbers will change the system behavior.
Further investigations certainly require experimental observation of this effect, and the settings as in Ref. \onlinecite{Pfeiffer08}, for instance, could be applied, which would be the subject of our future studies.

Annular Josephson junctions posses an enormous potential for various technological applications.
The fluxon dynamics as well as resonance phenomena are in the core of some of the most advanced ideas in superconducting digital technologies~\cite{Clarke, WallNat, Fed, Herr11, Mukhanov11, Volkmann13, Takeuchi}.
Another interesting application of annular Josephson junctions is in superconducting metamaterials~\cite{Kis}, which
generic element is a superconducting ring split by a Josephson junction.
One of the most recent studies have been dedicated to the resonant response of such metamaterials to the external signal in strongly nonlinear regimes~\cite{Kis}.
Regardless of the field in which the annular Josephson junctions have application, a good theoretical guideline and their understanding are crucial.
We hope that this work contributes to that understanding and that it will motivate further theoretical and experimental studies.

%============================================================================================================================================

\begin{acknowledgments}
	
J. Teki\' c whish tothank to Prof. Yu. M. Shukrinov and the BLTP, JINR, Dubna in Russia for their generous hospitality where a part of this work was done.
This work was supported by the grant of Fund for developing theoretical physics and mathematics "Bazis".
The reported study was funded by the RFBR
research projects  18-02-00318, 18-52-45011-IND.  The numerical simulations are supported by the RSF in the framework  of research project 18-71-10095.
This work was supported by the Serbian Ministry of
Education and Science under Contracts No. OI-171009 and No. III-45010
and  by the Provincial Secretariat for High Education and Scientific Research of Vojvodina (Project No. APV 114-451-2201).

\end{acknowledgments}

%==================================================================================================================

%===============================================================================================================


\begin{references}

\bibitem{Clarke}
J. Clarke,
Nature {\bf 425}, 133 (2003).

\bibitem{WallNat}
A. Wallraff, A. Lukashenko, J. Lisenfeld, A. Kemp, M. V. Fistul, Y. Koval, and A. V. Ustinov,
Nature {\bf 425}, 155 (2003).

\bibitem{Fed}
K. G. Fedorov, A. V. Shcherbakova,  M. J. Wolf, D. Beckmann, and A, V. Ustinov,
\prl {\bf 112}, 160502 (2014).

\bibitem{Herr11}
Q. P. Herr, A. Y. Herr, O. T. Oberg, and A.G. Ioannidis,
J. Appl. Phys. {\bf 109}, 103903 (2011).

\bibitem{Mukhanov11}
O. A. Mukhanov, IEEE Trans. Appl. Supercond. {\bf 21}, 760 (2011).

\bibitem{Volkmann13}
M. H. Volkmann, A. Sahu, C.J. Fourie, and O. A.  Mukhanov,
Supercond. Sci. Technol. {\bf 26}, 015002 (2013).

\bibitem{Takeuchi}
N. Takeuchi, Y. Yamanashi, and N. Yoshikawa,
Supercond. Sci. Technol. {\bf 28}, 015003 (2015).

\bibitem{Mazo}
J. J. Mazo and A. V. Ustinov
{\it The sine-Gordon model and its Applications}.
(Springer, Switzerland, 2014).

\bibitem{OBBook}
O. Braun and Yu.\ S.\ Kivshar,
{\it The Frenkel-Kontorova Model} (Springer, Berlin, 2003).

\bibitem{ACFK}
J. Teki\' c and P. Mali, {\it The ac driven Frenkel - Kontorova model}, (University of Novi Sad, Novi Sad, 2015).

\bibitem{Fult}
T. A. Fulton and R. C. Dynes,
Solid State Comm., {\bf 12}, 57 (1973).

\bibitem{Lucci15}
M. Lucci, D. Badoni, V. Merlo, I. Ottaviani, G. Salina, M. Cirillo, A. V. Ustinov, and D. Winkler,
\prl {\bf 115}, 107002 (2015).

\bibitem{DavPRL85}
A. Davidson, B. Dueholm, B. Kryger, and N. F. Pedersen
\prl  {\bf 55}, 2059 (1985).

\bibitem{Pfeiffer08}
J. Pfeiffer J., A. A. Abdumalikov, Jr. M. Schuster, and A. V. Ustinov,
\prb {\bf 77}, 024511 (2008).

\bibitem{Ustinov93}
A. V. Ustinov, M. Cirillo, and B. A. Malomed,
\prb {\bf 47}, 8357 (1993).

\bibitem{UstPRL}
A. V. Ustinov, T. Doderer, R. P. Huebener, N. F. Pedersen,
B. Mayer, and V. A. Oboznov
\prl {\bf 69}, 1815 (1992).

\bibitem{Zant95}
H. S. J. van der Zant, T. P. Orlando, S. Watanabe, and S. H. Strogatz,
\prl {\bf 74}, 174 (1995).

\bibitem{Wall}
A. Wallraff, A. V. Ustinov, V. V. Kurin, I. A. Shereshevsky, and N. K. Vdovicheva,
\prl {\bf 84}, 152 (2000).

\bibitem{Watanabe}
S. Watanabe, S. H. Strogatz, H. S. J. van der Zant, and T. P. Orlando,
\prl {\bf 74}, 379 (1995).

\bibitem{Nappi}
C. Nappi, M. P. Lisitskiy, G. Rotoli, R. Cristiano, and A. Barone,
\prl {\bf 93}, 187001 (2004).

\bibitem{Monaco}
R. Monaco, J. Mygind, and V. P. Koshelets,
\prb {\bf 100}, 064501 (2019).

\bibitem{Lomdhal}
P. S. Lomdhal, O. H. Soerensen, and P. L. Christiansen,
Physica Scripta, {\bf 25}, 879 (1982).

\bibitem{Kawamoto}
H. Kawamoto,
Prog. Theor. Phys., {\bf 70}, 1171 (1983).

\bibitem{ShapPRL}
S. Shapiro, \prl {\bf 11}, 80 (1963).

\bibitem{UstPD}
A. V. Ustinov,
Physica D {\bf 123}, 3315 (1998).

\bibitem{ShukLNCS2012} Yu. Shukrinov, I. Rahmonov, M. Hamdipour,
Lecture Notes in Computer Science {\bf 7125}, 234 (2012).

\bibitem{ShukJETP2012} Yu. M. Shukrinov, I. R. Rahmonov,
JETP {\bf 115}, 289 (2012).

\bibitem{Kis}
E. I. Kiselev, A. S. Averkin, M. V. Fistul, V. P. Koshelets, and A. V. Ustinov,
arXiv:1905.01511v3 [physics.app-ph] (2019).

\end{references}
\end{document}